\documentclass[aps,prl,twocolumn,superscriptaddress,draft]{revtex4}
\input epsf

\newcommand{\eg}{{\em e.g.\ }}

\newcommand{\ra}{\rightarrow}

\newcommand{\lamk}{$\Omega^- \ra \Lambda K^-, \; \Lambda \ra p \pi^-$}
\newcommand{\lampi}{$\Omega^- \ra \Lambda \pi^-, \; \Lambda \ra p \pi^-$}
\newcommand{\omlampi}{$\Omega^- \ra \Lambda \pi^-$}
\newcommand{\xipi}{$\Omega^- \ra \Xi^0 \pi^-, \; \Xi^0 \ra p \pi^-$}
\newcommand{\xippi}{$\Xi^0 \ra p \pi^-$}

\begin{document}

\title{Search for $\Delta$S = 2 Nonleptonic Hyperon Decays}

\affiliation
{Institute of Physics, Academica Sinica, Taipei 11529, Taiwan, Republic of China}
\affiliation{University of California, Berkeley, California 94720, USA}
\affiliation{Fermi National Accelerator Laboratory, Batavia, Illinois 60510, USA}
\affiliation{University of Guanajuato, 3700 Leon, Mexico}
\affiliation{Illinois Institute of Technology, Chicago, Illinois 60616, USA}
\affiliation{University of Lausanne, CH-1015 Lausanne, Switzerland}
\affiliation{Lawrence Berkeley National Laboratory, Berkeley, California 94720, USA}
\affiliation{University of Michigan, Ann Arbor, Michigan 48109, USA}
\affiliation{University of South Alabama, Mobile, Alabama, 36688, USA}
\affiliation{University of Virginia, Charlottesville, Virginia 22904, USA}
\author{C.~G.~White}
\email[corresponding author: ]{whitec@iit.edu}
\affiliation{Illinois Institute of Technology, Chicago, Illinois 60616, USA}
\author{R.~A.~Burnstein}
\author{A.~Chakravorty}
\affiliation{Illinois Institute of Technology, Chicago, Illinois 60616, USA}
\author{A.~Chan}
\author{Y.~C.~Chen}
\affiliation
{Institute of Physics, Academica Sinica, Taipei 11529, Taiwan, Republic of China}
\author{W.~S.~Choong}
\affiliation{University of California, Berkeley, California 94720, USA}
\affiliation{Lawrence Berkeley National Laboratory, Berkeley, California 94720, USA}
\author{K.~Clark}
\affiliation{University of South Alabama, Mobile, Alabama, 36688, USA}
\author{E.~C.~Dukes}
\author{C.~Durandet}
\affiliation{University of Virginia, Charlottesville, Virginia 22904, USA}
\author{J.~Felix}
\affiliation{University of Guanajuato, 3700 Leon, Mexico}
\author{G.~Gidal}
\author{P.~Gu}
\affiliation{Lawrence Berkeley National Laboratory, Berkeley, California 94720, USA}
\author{H.~R.~Gustafson}
\affiliation{University of Michigan, Ann Arbor, Michigan 48109, USA}
\author{C.~Ho}
\affiliation
{Institute of Physics, Academica Sinica, Taipei 11529, Taiwan, Republic of China}
\author{T.~Holmstrom}
\author{M.~Huang}
\affiliation{University of Virginia, Charlottesville, Virginia 22904, USA}
\author{C.~James}
\affiliation{Fermi National Accelerator Laboratory, Batavia, Illinois 60510, USA}
\author{C.~M.~Jenkins}
\affiliation{University of South Alabama, Mobile, Alabama, 36688, USA}
\author{D.~M.~Kaplan}
\author{L.~M.~Lederman}
\affiliation{Illinois Institute of Technology, Chicago, Illinois 60616, USA}
\author{N.~Leros}
\affiliation{University of Lausanne, CH-1015 Lausanne, Switzerland}
\author{M.~J.~Longo}
\author{F.~Lopez}
\affiliation{University of Michigan, Ann Arbor, Michigan 48109, USA}
\author{L.~C.~Lu}
\affiliation{University of Virginia, Charlottesville, Virginia 22904, USA}
\author{W.~Luebke}
\affiliation{Illinois Institute of Technology, Chicago, Illinois 60616, USA}
\author{K.~B.~Luk}
\affiliation{University of California, Berkeley, California 94720, USA}
\affiliation{Lawrence Berkeley National Laboratory, Berkeley, California 94720, USA}
\author{K.~S.~Nelson}
\affiliation{University of Virginia, Charlottesville, Virginia 22904, USA}
\author{H.~K.~Park}
\affiliation{University of Michigan, Ann Arbor, Michigan 48109, USA}
\author{J.-P.~Perroud}
\affiliation{University of Lausanne, CH-1015 Lausanne, Switzerland}
\author{D.~Rajaram}
\author{H.~A.~Rubin}
\affiliation{Illinois Institute of Technology, Chicago, Illinois 60616, USA}
\author{P.~K.~Teng}
\affiliation
{Institute of Physics, Academica Sinica, Taipei 11529, Taiwan, Republic of China}
\author{J.~Volk}
\affiliation{Fermi National Accelerator Laboratory, Batavia, Illinois 60510, USA}
\author{S.~L.~White}
\affiliation{Illinois Institute of Technology, Chicago, Illinois 60616, USA}
\author{P.~Zyla}
\affiliation{Lawrence Berkeley National Laboratory, Berkeley, California 94720, USA}
\collaboration{HyperCP Collaboration}
\noaffiliation

\date{\today}

\begin{abstract}
A sensitive search for the rare decays $\Omega^- \ra \Lambda \pi^-$
and $\Xi^0 \ra p \pi^-$ has been performed using data from the 1997
run of the HyperCP (Fermilab E871) experiment. Limits on other such processes do not
exclude the possibility of observable rates for $|\Delta S| = 2$
nonleptonic hyperon decays, provided the decays occur through
parity-odd operators. We obtain the branching-fraction limits
$\mathcal{B}($\omlampi$)< 2.9\times 10^{-6}$ and $\mathcal{B}($\xippi$)< 8.2\times
10^{-6}$, both at 90\% confidence level.
\end{abstract}


\maketitle


The standard model allows $|\Delta S| = 2$ transitions through
second-order weak interactions.  This approach successfully
describes $K^0\overline{K}{}^0$ mixing, which is currently the only
observed $|\Delta S| = 2$ transition.  While other $|\Delta S| = 2$
processes have generally been considered too highly suppressed to be
observed experimentally, it has been noted that the rate of
$K^0\overline{K}{}^0$ mixing does not exclude nonleptonic $|\Delta S|
= 2$ hyperon decays at observable rates, provided that they proceed
through new parity-odd channels~\cite{Glashow,valencia}.
Measurements can thus be used to set limits on parity-odd
contributions to hyperon decays.  It is of interest, therefore, to
perform sensitive searches for such decays. There is also interest
in searches for direct $|\Delta S| = 2$ transitions in $B$-meson
decays~\cite{Singer}.

Observation of $|\Delta S| = 2$ nonleptonic hyperon decays at
current levels of sensitivity would strongly suggest new physics.
He and Valencia~\cite{valencia} have
parametrized the strength of any new parity-odd interaction as a
ratio to the strength of the electroweak interaction. This ratio is
constrained by hyperon branching ratios~\cite{valencia}, \eg,
\begin{equation}
\mathcal{B}(\Xi^0 \rightarrow p \pi^-) = 0.9\left(\frac{\alpha_{\rm new}}{\alpha_{\rm EW}}\right)^2\,,
\end{equation}

\noindent where $\alpha_{\rm new}$ ($\alpha_{\rm EW}$) is the strength of the new (electroweak)
interaction. Current experimental limits on $|\Delta \mbox{S}| = 2$
decays~\cite{PDG} include $\mathcal{B}(\Omega^- \rightarrow \Lambda \pi^-$)
$\leq1.9 \times 10^{-4}$~\cite{Bourquin} and $\mathcal{B}$(\xippi) $\leq 3.6
\times 10^{-5}$~\cite{Geweniger}, both at 90\% confidence level
(C.L.). We report a search for these decays using data from the 1997
run of HyperCP (Fermilab Experiment 871). These data are well suited
for such studies as they contain large numbers of charged hyperon
decays, $\sim$$10^9$ $\Xi^-\to\Lambda\pi^-$ and $\sim$
$10^7$ $\Omega^-\to\Lambda K^-$ decays.

\begin{figure}
\centerline{\epsfxsize=3.4 in \epsffile {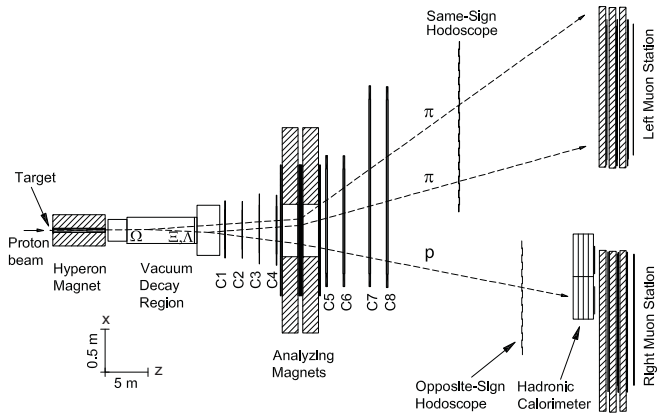}}
\caption{Plan view of the HyperCP spectrometer.  Note that the {\em z}
scale is compressed by a factor of 10 with respect to the {\em x} scale. \label{fig:HyperCP}}
\end{figure}

The HyperCP spectrometer is shown schematically in
Fig.~\ref{fig:HyperCP}. The spectrometer (described in detail
elsewhere~\cite{spectrometer}) was designed to have large acceptance
for the decay chain $\Xi^- \rightarrow \Lambda \pi^-, \, \Lambda
\rightarrow p \pi^-$. In brief, a negatively charged secondary beam
was formed by the interaction of $800\;$GeV/c primary protons from the
Tevatron in a $0.2\times 0.2\times 6\,{\rm cm}^3$ copper target,
with the secondaries sign- and momentum-selected by means of a
6.096-m-long curved collimator within a 1.667~T dipole magnetic
field (Hyperon Magnet).
The mean secondary momentum was about 160\,GeV/$c$ with momenta
ranging from 120 to 250\,GeV/$c$. Hyperon decays occurring within a
13-m-long evacuated decay pipe were reconstructed in three
dimensions using a series of high-rate, narrow-gap multiwire proportional
chambers arrayed on either side of a pair of
dipole magnets (Analyzing Magnets).

The trigger for the data acquisition system~\cite{DAQ} used
scintillation-counter hodoscopes located sufficiently far downstream
of the Analyzing Magnets so that the hyperon decay products were
well separated from the secondary beam. A coincidence was required
of at least one (``same-sign") hodoscope hit consistent with a charged particle of
the same sign as the secondary beam and at least one
``opposite-sign" hit. 
To suppress muon and low-energy backgrounds the
trigger also required a minimum energy deposit in the Hadronic
Calorimeter.

\begin{figure}
\centerline{\epsfxsize=3. in \epsffile {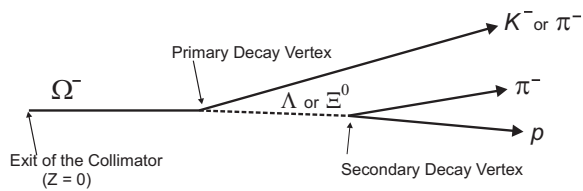}}
\caption{Event topology for all decays considered here.\label{fig:gfit}}
\end{figure}

\begin{figure}
\centerline{\epsfxsize=3.4 in \epsffile {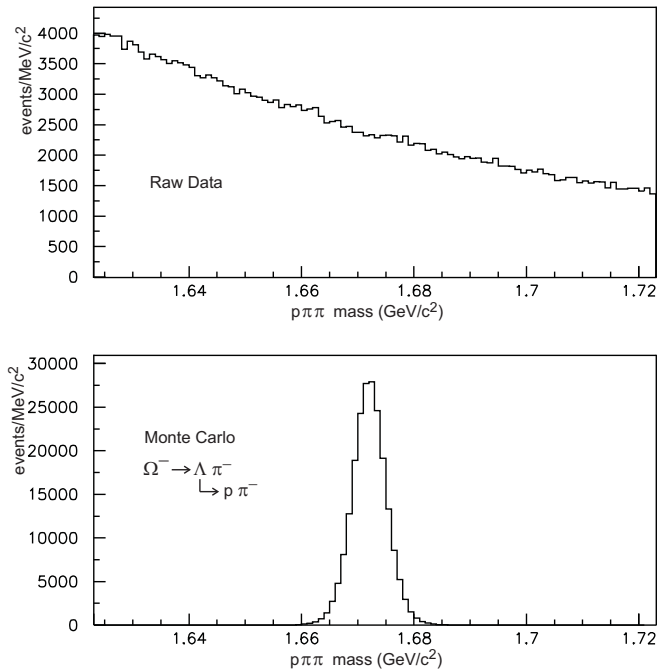}}
\caption{The $p\pi^-\pi^-$ invariant-mass distribution prior to event selection (top) and from a Monte Carlo
simulation of the $\Omega^-\to\Lambda\pi^-$, $\Lambda\to p\pi^-$ process (bottom).\label{fig:pith}}
\end{figure}

We searched for events consistent with either $\Delta S = 2$ decay chain,
$\Omega^-\to\Lambda\pi^-$, $\Lambda\to p\pi^-$ or
$\Omega^-\to\Xi^0\pi^-$, $\Xi^0\to p\pi^-$.  Also studied were the
copious events from the $\Omega^-\to\Lambda K^-$, $\Lambda\to
p\pi^-$ decay sequence used for normalization. Such events all have
the topology shown in Fig.~\ref{fig:gfit}, with three charged tracks
forming two separated vertices.  A least-squares geometric fit
determined the positions of the primary and secondary vertices, as
well as the chi-square ($\chi^2$) value for the event to have the required topology.
The reconstructed $\Omega^-$ trajectory was traced back through the
Hyperon Magnet using the measured magnetic field to determine its $x$
and $y$ coordinates at the midpoint of the target.
Figure~\ref{fig:pith} compares the $p\pi^-\pi^-$ invariant-mass
distribution of all events before event selection with that of Monte
Carlo-generated $\Omega^-\to\Lambda\pi^-$, $\Lambda\to p\pi^-$
events.

Selection criteria for signal events were based on Monte Carlo
simulations and studies of a purified sample of $\Omega^-\to\Lambda
K^-$, $\Lambda\to p\pi^-$ events. Events were required to have three
charged tracks with two tracks on the ``same-sign" side of the
spectrometer and one on the ``opposite-sign" side.  Most $K^- \ra
\pi^-\pi^-\pi^+$ events were excluded by requiring the invariant
mass (treating all three charged tracks as pions) to exceed
0.500\,GeV/$c^2$, 3 standard deviations ($\sigma$) above
the $K^-$ mass~\cite{PDG}. Since many background
events originated from secondary interactions near the exit of the
collimator or at the exit windows of the Vacuum Decay Region, all
events were required to have a primary-vertex $z$ position between
150 and 1180\,cm from the exit of the collimator, well within the Vacuum Decay Region.  The secondary
vertex was required to be downstream of the primary vertex.  The
$\chi^2/$degree-of-freedom from the geometric fit was required to be
less than 1.7; this requirement was 98\% efficient for
$\Omega^-\to\Lambda K^-$, $\Lambda\to p\pi^-$ events.  Selection
criteria for the projected $\Omega^-$ position at the target were
determined using the purified $\Omega^-\to\Lambda K^-$, $\Lambda\to
p\pi^-$ sample.  Due to differing resolutions in $x$ and $y$ at the
target, an elliptical target cut was used which was 92\% efficient for
$\Omega^-\to\Lambda K^-$, $\Lambda\to p\pi^-$.

\begin{figure}
\centerline{\epsfxsize=3.4 in \epsffile {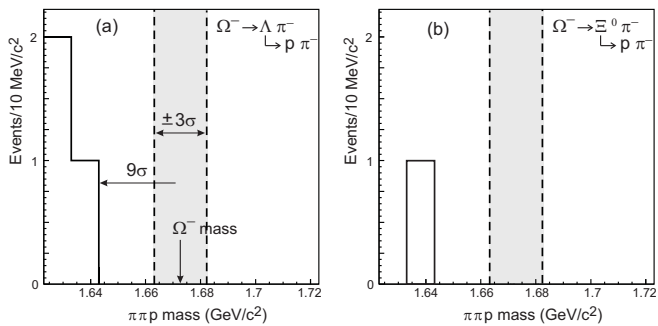}}
\caption{The $p\pi^-\pi^-$ invariant-mass distributions for events satisfying all selection criteria for
(a) $\Omega^-\to\Lambda\pi^-$, $\Lambda\to p\pi^-$ and (b) $\Omega^-\to\Xi^0\pi^-$, $\Xi^0\to p\pi^-$.\label{fig:data}}
\end{figure}

Additional selection criteria were specific to each mode. For the
$\Omega^-\to\Lambda\pi^-$, $\Lambda\to p\pi^-$ search we required
the invariant mass of the $p\pi$ combination forming the secondary
vertex to be within $\pm$2.0\,MeV/$c^2$ ($\pm 1.5$ times the
rms mass resolution) of the $\Lambda$ mass~\cite{PDG}.
Decay polar angles in the respective hyperon rest frames were calculated
with respect to the parent's lab-frame momentum vector.
Background events tended to have a small decay polar angle for the pion in the $\Lambda\pi^-$
center-of-mass system ($\theta_\pi<0.82$\,rad), thus we required $\theta_\pi>0.82$\,rad.
The invariant-mass distribution, assuming a $p\pi^-\pi^-$ final
state, after all selection cuts is shown in Fig.~\ref{fig:data}a. No
events were observed within $\pm9\,\sigma$ of the expected mass.

For the $\Omega^-\to\Xi^0\pi^-$, $\Xi^0\to p\pi^-$ search, we
required that the proton momentum be more than 38\% of the
$p\pi^-\pi^-$ total momentum, that the invariant mass of the
$p\pi$ combination forming the secondary vertex be within
$\pm5.0\,$MeV/$c^2$ ($\pm 3.0\,\sigma$) of the $\Xi^0$
mass~\cite{PDG}, that the $z$ position of the daughter-hyperon decay
vertex be within the Vacuum Decay Region, and that the polar angle of the proton
in the $p\pi$ center-of-mass frame be less than $2.97$\,rad.  The
$p\pi^-\pi^-$ invariant-mass distribution for events satisfying all
selection criteria is shown in Fig.~\ref{fig:data}b.  Again, no
events were observed within $\pm9\sigma$ of the expected mass.

The spectrometer acceptances for these decays were estimated using a
Monte Carlo simulation. The generated
$\Omega^-$ momentum and position distributions at the target were
tuned to match those observed in the data for a purified sample of
$\Omega^-\to\Lambda K^-$, $\Lambda\to p\pi^-$ decays.
For all decay modes, the acceptance includes the probability that
the parent decay occur within the Vacuum Decay Region.
The geometric acceptance of the spectrometer was determined
primarily by the apertures of the Analyzing Magnets. Due to the
larger $Q$ values for $\Omega^-\to\Lambda\pi^-$ and $\Xi^0\to
p\pi^-$ compared to those of $\Omega^-\to\Lambda K^-$ and
$\Lambda\to p\pi^-$, tracks from signal-mode decays would be
approximately three times more likely to miss the magnet apertures and
be lost.  For events accepted in the magnet apertures, the trigger
efficiency ranged from 93\% for \lampi$\;$ candidates to 99\% for \lamk$\;$
events. The candidate signal- and normalizing-mode events were all
accepted by the same trigger, so it was not necessary to
cross-calibrate trigger efficiencies by mode. Offline
event-selection efficiencies differed considerably by mode, since
restrictive selection criteria to suppress background were required
for the signal modes but not for the normalizing mode. To study the
systematic uncertainties of the selection efficiencies, the
selection criteria were varied. This produced only slight changes in the
results.


The $\alpha$ decay parameter for $\Omega^- \to \Lambda K^-$ has
recently been precisely measured and found to be small, but most likely
not zero~\cite{Lanchun}, while the measured value for $\Omega^- \to
\Xi^0 \pi^-$ is less precise and is consistent with zero~\cite{PDG}.
Theory predicts little parity violation in the dominant $\Omega^-$
decays~\cite{theory} (as observed); however, we cannot
exclude large decay asymmetries for $\Omega^- \to \Lambda \pi^-$.
Similarly, we do not know the size of parity violation in $\Xi^0 \to
p \pi^-$. We therefore assign zero to the value of $\alpha$
for $\Omega^- \to \Lambda \pi^-$ and $\Xi^0 \to p \pi^-$. The
dependence of the acceptance on $\alpha$ is tabulated in
Table~\ref{tab:syst}. The acceptance times the selection
efficiency (including the maximal acceptance variation due to
uncertainty in $\alpha$) was ${6.2}^{+0.9}_{-1.0}$\% for \xipi,
${6.9}^{+1.8}_{-2.2}$\% for $\Omega^-\to\Lambda\pi^-$, $\Lambda\to
p\pi^-$, and ($34.9\pm1.2$)\% for \lamk.

 \begin{table}
 \caption{Relative spectrometer acceptances vs. $\alpha$ decay parameters.
 \label{tab:syst}}
 \begin{tabular}{r@{.}lcccr@{.}lr@{.}lc}\hline\hline
 \multicolumn{4}{c}{$\Omega^- \to \Lambda \pi^-;\,\, \Lambda \to p \pi^-$} & $\;\;\;\;\;\;$ &
 \multicolumn{5}{c}{$\Omega^- \to \Xi^0 \pi^-;\,\, \Xi^0 \to p \pi^-$} \\
 \multicolumn{2}{c}{$\alpha_{\Omega^-}$}&$\alpha_{\Lambda}$&Rel.
 Acc.&&\multicolumn{2}{c}{$\alpha_{\Omega^-}$}&\multicolumn{2}{c}{$\alpha_{\Xi^0}$}&Rel. Acc.\\ \hline
 $-$1 & $00\;\;\;\;$ & 0.642 & 1.17 && $-$0 & 05$\;\;\;$ & $-$1&0 & 1.05 \\
 $-$0 & 75       & 0.642 & 1.13 && $-$0 & 05 & 0   &0 & 1.05 \\
 $-$0 & 50       & 0.642 & 1.09 && $-$0 & 05 & 1   &0 & 1.04 \\
 $-$0 & 25       & 0.642 & 1.05 && 0    & 09 & $-$1&0 & 0.98 \\
 0    & 00       & 0.642 & 1.00 && 0    & 09 & 0   &0 & 1.00 \\
 0    & 25       & 0.642 & 0.94 && 0    & 09 & 1   &0 & 1.02 \\
 0    & 50       & 0.642 & 0.87 && 0    & 23 & $-$1&0 & 0.92 \\
 0    & 75       & 0.642 & 0.79 && 0    & 23 & 0   &0 & 0.95 \\
 1    & 00       & 0.642 & 0.70 && 0    & 23 & 1   &0 & 0.98 \\\hline\hline
 \end{tabular}
 \end{table}

\begin{figure}
\centerline{\epsfxsize=3.4 in \epsffile {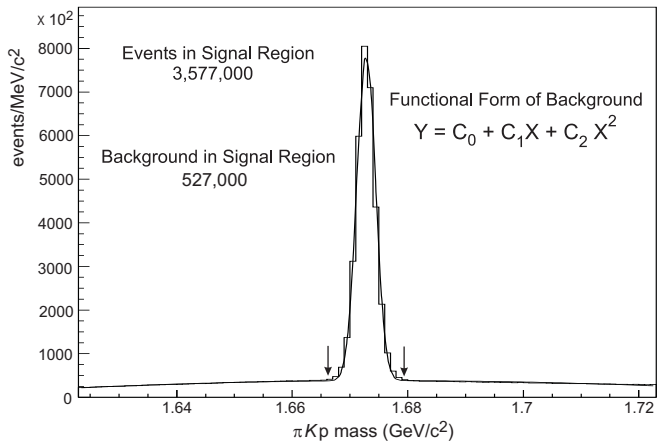}}
\caption{Observed $pK^-\pi^-$ invariant-mass distribution.
The number of events in the $\Omega^-\to\Lambda K^-$, $\Lambda\to p\pi^-$
peak was used to normalize the upper limits presented here. The arrows indicate $\pm 3 \sigma$ in mass resolution. \label{fig:norm}}
\end{figure}

Signal-mode branching ratios were normalized using the observed
$\Omega^-\to\Lambda K^-$, $\Lambda\to p\pi^-$ sample, whose invariant-mass distribution is shown in
Fig.~\ref{fig:norm}. The figure includes a fit of the mass
distribution to a Gaussian plus a second-order polynomial. The
number of normalizing-mode decays was estimated using three
different background-subtraction methods. In all three, an estimate
of the background was subtracted from the total number of events
observed within $\pm3\,\sigma$ of the $\Omega$ mass. The first
background estimate was the sum of all events within $\pm3\,\sigma$ of
the $\Omega$ mass minus the integral of the fitted second-order
polynomial over that region. In the second method, a fit was
performed over a limited region ($\pm 3\,\sigma$) around the $\Omega$
mass to a Gaussian plus a constant. The third method averaged the
two bins at $\pm 3\,\sigma$ and multiplied this by
the number of bins within the $\pm3\,\sigma$ window.  All methods gave
similar results.  The final estimate of the number of observed
normalizing events was taken as the average of the three estimates,
$(3.050 \pm 0.023)\times10^6$, with the uncertainty defined as half
the difference between the largest and smallest estimates
(corresponding to about $1 \,\sigma$, assuming Gaussian statistics).
After correcting for acceptance and selection efficiencies, and
accounting for the measured branching fractions, the total number of
$\Omega^-$ baryons exiting the collimator was $(20.2 \pm 0.8)
\times 10^6$.

Signal-mode branching fractions were obtained from
\begin{eqnarray}
\mathcal{B}(\Omega^-\to\Lambda\pi^-)&=&\frac{N_{sig}}{N_{norm}}\frac{A_{norm}}{A_{sig}}\mathcal{B}(\Omega^-\to\Lambda K^-)\,,\label{eq:lampi}\\
\mathcal{B}(\Xi^0\to p\pi^-)&=&\frac{N_{sig}}{N_{norm}}\frac{A_{norm}}{A_{sig}} \nonumber \\
&\times&\frac{\mathcal{B}(\Omega^-\to\Lambda K^-)\mathcal{B}(\Lambda\to p\pi^-)}{\mathcal{B}(\Omega^-\to\Xi^0\pi^-)}\,.\label{eq:ppi}
\end{eqnarray}
Here $N$ denotes the number of events observed and $A$ the
acceptance for a given mode, with subscripts $sig$ designating the
signal mode in question and $norm$ the normalizing mode
$\Omega^-\to\Lambda K^-,\Lambda\to p\pi^-$. The number of signal
events observed is zero in each case. The measured branching ratios
entering into Eqs.~(\ref{eq:lampi}) and (\ref{eq:ppi}) are
$\mathcal{B}(\Omega^-\to\Lambda K^-)=(67.8\pm 0.7)$\%, $\mathcal{B}(\Lambda\to
p\pi^-)=(63.9\pm 0.5)$\%, and
$\mathcal{B}(\Omega^-\to\Xi^0\pi^-)=(23.6\pm0.7)$\%~\cite{PDG}. Systematic
uncertainties include the uncertainties of these branching ratios,
that of the normalizing-mode background subtraction, and, most
importantly, those associated with acceptances and event selection.

To derive 90\%-C.L. upper limits for the numbers of events observed
($U_n$), we included systematic uncertainties~\cite{Poisson} as
follows:
\begin{equation}
U_n = U_{n0}[1+(U_{n0}-n)\sigma^{2}_r/2]\,.
\label{eq:syst}
\end{equation}
Here $U_{n0}$ represents the statistical limit based on the Poisson
distribution with no systematic uncertainties, $n$ is the number of
observed events, and $\sigma_r$ is the relative systematic
uncertainty.  In our case ($n=0$), Eq.~(\ref{eq:syst})
reduces to
\begin{equation}
U_0 = 2.3(1+2.3\sigma^{2}_r/2)\,.\label{eq:syst1}
\end{equation}


The relative uncertainty (including uncertainties in the acceptance, selection, normalization, and branching ratios) was
16\% for \xipi and 32\% for \lampi. Upper limits on the numbers of observed events at the 90\% C.L.
thus determined are 2.4 events for \xippi and 2.6 for \omlampi,
comparable to the 2.3 events obtained from a frequentist statistical
treatment of the Poisson fluctuation alone~\cite{PDG}.
We thus obtain $\mathcal{B}($\omlampi$)< 2.9\times 10^{-6}$ and $\mathcal{B}($\xippi$)<
8.2\times 10^{-6}$, both at 90\% C.L. These results represent
improvements by one to two orders of magnitude over previous
measurements.

\begin{acknowledgments}
We thank the staffs of Fermi National Accelerator Laboratory and
each collaborating institution for support necessary to
complete these studies.  This work was supported by the
US Department of Energy, the National Science Council of Taiwan,
ROC, and the Swiss National Science Foundation. E$.$C$.$D.\
and K$.$S$.$N.\ were partially supported by the Institute
for Nuclear and Particle Physics. K$.$B$.$L.\ was partially
supported by the Miller Institute for Basic Research in Science. D$.$M$.$K.\
acknowledges support from the Particle Physics and Astronomy
Research Council of the United Kingdom and the hospitality of
Imperial College London while this paper was in preparation.
\end{acknowledgments}

\end{document}